\def\be            {\begin{eqnarray}}

\def\ee            {\end{eqnarray}}

\newcommand\erf[1] {(\ref{#1})}

\def\ii            {{\rm i}}

\newcommand\void[1]{}
\def\zet           {\ensuremath{\mathbb Z}}

\documentclass[notoc]{JHEP3} 
\usepackage{latexsym,amsmath,amssymb,amsfonts}
\usepackage[mathscr]{eucal}
\numberwithin{equation}{section}
\def\hbar          {\ensuremath{{\liefont h}}}

\title{D-branes in the diagonal SU(2) coset}
\author{Albrecht Wurtz \\
Karlstads Universitet\\ 
Universitetsgatan 5 \\
S\,--\,651\,88\, Karlstad\\
Email: \email{albrecht.wurtz@kau.se}
}

\abstract{ 
The symmetry preserving D-branes in coset theories 
have previously been described
as being centered around projections of 
products of conjugacy classes in the underlying Lie groups. 
Here, we investigate the coset where a diagonal action of 
$SU(2)$ is divided out from $SU(2)\times SU(2)$. 
The corresponding target space is described as a (3-dimensional) 
pillow with four distinguished corners. It is shown that the (fractional) brane which corresponds to the fixed point that
arises in the CFT description, is spacefilling. 
Moreover, the spacefilling brane
is the only one that reaches all of the corners. 
The other branes are 3, 1 and 0 - dimensional.}


\preprint{}

\begin{document}

\section{Introduction}
In a Lagrangian description of CFT, 
one must impose boundary conditions for
open world-sheets. These boundary conditions should be such   
that there is an action of the chiral symmetry algebra
on open world-sheet correlators, in particular the resulting boundary CFT
should be conformal.
In free bosonic theories, this is achieved by simply imposing 
Neumann- or Dirichlet boundary conditions, cf.\ \cite{Polchinski}.
In curved backgrounds, the situation is more complicated.
For instance in WZW models (describing strings propagating in Lie groups), 
the symmetry preserving D-branes
are localized at certain conjugacy classes, cf.\ \cite{GawG}
\cite{alsc2}. 
\\[-3mm]

On the other hand, one can describe D-branes from a perspective
which is more oriented towards the algebraic description of CFT on the world-sheet.
In that approach, one {\it begins} with imposing conformal invariance
and other requirements on so-called boundary states \cite{Cardy}.
Then, one can analyze the shapes of these boundary states. In
free bosonic theories there is full agreement between
these descriptions \cite{DiV},
but already in WZW models, 
the agreement is less direct \cite{FFFS}.\\[-3mm] 

A class of CFT models related to WZW models 
are the coset models. The relevant target spaces are adjoint quotients
$G/{\rm Ad }(H)$, where $H\subset G$ is an embedded Lie subgroup of 
the connected, simply connected Lie group $G$, cf. \cite{GaKu} \cite{schnitzer}. 
The D-branes in the Lagrangian
description are certain projected products of conjugacy classes \cite{Ga}. 
In general, the CFT description of coset models 
involve a certain identification group \cite{scya5} \cite{ffrs2} \cite{fusS4}.
In most of the literature on D-branes in such models, 
only situations where the identification group acts freely are considered, 
i.\ e.\ $\mathcal G _{\rm id}$ has no fixed points. 
The aim of the present note is to perform a geometric analysis
of a model where such fixed points do occur, 
with interest primarily on the branes that correspond to the fixed point.
We consider the sigma model with
target space the coset \\[-6mm]
\be Q &=& \frac{SU(2)\times SU(2)}{{\rm Ad}(SU(2))} \,,
\label{pillow}
\ee 
which is the set of equivalence classes $[ g_1,g_2 ]$
with the equivalence relation 
\be (g_1,g_2) \sim (g g_1 g^{-1} , gg_2 g^{-1})\,,
\label{ca}\ee
and with $g,g_1,g_2 \in SU(2)$. 
The sigma model on $Q$ can be described as a gauged WZW model \cite{Ga},
which after choosing levels $k,l,k+l\in \zet_{>0}$ for the three occuring factors,
corresponds to a coset CFT. 
The particular choice $l=1$ gives the series of Virasoro minimal models. 
In the sequel, we shall think of both $l$ and $k$ as being large, so that we can make use of the existing results on boundary conditions in coset models 
\cite{Ga}, \cite{FS} at large level. \\[-3mm]

In the next section, the geometry of the target space is described. 
The space $Q$ is a pillow, or a pinched cylinder, with four distinguished corners. 
We introduce coordinates and classify degenerate points, for which the stabilizer
of the coset action \erf{ca} is larger than at generic points. In the third and final
section, the geometry of the subsets where the branes are centered is discussed. The 
dimensionalities of the branes are derived, 
and the branes reaching the corners of $Q$ are classified.


\section{Target space geometry}
An adjoint diagonal coset \erf{pillow} can be parametrized as follows. The set $Q$ consists of equivalence classes
\be [g_1,g_2]=[gg_1g^{-1},gg_2g^{-1}]~~~~~~g,g_1,g_2\in G\,.
\ee
For any $[g_1,g_2]$ we can choose a representative of $g_1\sim t_1$ in $T_W=T/W$,
where $T$ is a maximal torus and $W$ is the Weyl group. 
It remains to determine where we can choose a representative of 
$g_2$ with the remaining 'gauge freedom' $S_{t_1}$. 
Henceforth, $G$ will denote $SU(2)$. 
The stabilizers at the center $Z=\{ \pm e \}$ are given by $S_{\pm e}=G$,
and for other $t_1\in T_W \slash Z$ we have $S_{t_1}=T$. This gives the fibration
\be Q = \left\{ (t,q) ~~|~~ t\in T_W~,~~~\begin{array}{ccc}
            q\in T_W~~  &  {\rm if}~~t=\pm e        \\[2mm]
            q\in G/{\rm Ad}(T)~~  & {\rm else}   
           \end{array}
\right\}\,.
\ee
Using the Pauli matrices $\sigma_i$ with relations 
$\sigma_i \sigma_j=\delta _{ij}\mathbf 1+\ii \epsilon_{ijk}\sigma_k$, 
one can parametrize group elements as
\be g(\psi,\theta,\phi) 
= \cos \psi \,\mathbf 1
+\ii \sin \psi \, \bar n \cdot \bar \sigma\,.
\label{notation}
\ee
Here $\bar n=\bar n(\theta,\phi)\in S^2$ is given by the standard spherical coordinates $\theta$ and $\phi$, and 
$T_W$ is the interval parametrized by $0\leq \psi\leq \pi$.
Note that we have 
\be g^{-1} = g(-\psi,\theta,\phi)=g(\psi,\pi{-}\theta,\pi{+}\phi)= g^\dagger \,.
\label{inverse}
\ee
The geometry of the parafermion coset $PF=G/{\rm Ad}(T)$ 
was described in \cite{MMS1}
as a round disc. It can also be described as disc with two `corners' \cite{FW}. 
Thus, we may think of $Q$ as a pillow, or as a pinched cylinder \cite{FS} 
where the corners are pairwise connected by a singular line.
The description above has an arbitrariness in that we could have equally well
fixed $\psi_2 \in T_W$. Then, we could have a pinched cylinder with the singular line
going between other pairs of corners, instead. 
Either way, \erf{pillow} has four corners. 
In the bulk, it is three-dimensional. We choose as coordinates the angles 
$\psi_1$, $\psi_2$ and $\theta$, where $\theta$ may be 
either $\theta_1$ or $\theta_2$. At the edges of the pillow, $\psi_1,\psi_2=0,\pi$, 
the coordinate $\theta$ has no meaning. The ambiguity in $\theta$ is the only ambiguity, 
we cannot get rid of $\psi_1$ 
and $\psi_2$ as coordinates (assuming a fixed choice of coordinates on the covering). 
In any case, the coordinates $\phi_1$ and $\phi_2$ are gauged away. 

\paragraph{Stabilizers} We can in a natural way think of $Q$ as embedded in its covering 
$G \times G$, via the above described coordinates. 
The stabilizer of the Ad$(G)$-action on an element 
$(t,q)\in Q \subset G \times G$ is given by those $g \in G$ for which
\be (t,q)=(gtg^{-1},gqg^{-1})\,, ~\Leftrightarrow~ g\in S_{(t,q)}= S_t\cap S_q \,.
\ee
For $t=\pm e$, $S_t = G$, else $S_t = T$. Since every element $q\in G$ can be brought to $T$ 
via conjugation; $cqc^{-1}=t_q\in T$ with an element $c\in G$,
it follows that the generic stabilizers are given by $S_q =c^{-1} T c $, with $c\in G$
a fixed element (depending on $q$). 
Assume $q\notin T$ and let $s\in S_{t_q}$, which equals $T$, then
\be (c^{-1}sc)q(c^{-1}sc)^{-1}  
=c^{-1} s t_q s^{-1} c = c^{-1} t_q c = q\,.
\ee
This shows $c^{-1} T c \subset S_q$. Likewise, one can show $S_q \subset c^{-1} T c$: 
Suppose $a\in S_q$, then 
\begin{eqnarray}aqa^{-1}=q  ~~
& \Rightarrow & ~~ caqa^{-1}c^{-1}=t_q  \nonumber \\
&\Rightarrow & ca(c^{-1}t_qc)a^{-1}c^{-1}=t_q \nonumber \\
&\Rightarrow & cac^{-1}\in S_{t_q}=T\,, \nonumber \end{eqnarray}
and it follows that $c^{-1} T c = S_q$. 
Indeed, one may choose $T$ to be any maximal torus, 
and one could equally well have defined $T = S_q $. 
Note that for $c\neq \pm e$, we have $c^{-1} T c \cap T = Z$, the center of the group. 
The analysis is summed up in the following table.

\begin{center}
\begin{tabular}{c|c|c|c|c|c}
$t\in$      & $q\in$      & $S_t$ & $S_q$     & $S_t \cap S_q$ & $(t,q)\in Q$ \\[1mm] \hline \\[-4mm] 
$Z$         & $ Z$        & $G$    & $G$        & $G$             & corner \\[3pt] 
            & $T{\setminus} Z$ & $G$    & $T$        & $T$             & edge/boundary \\[3pt] 
$T{\setminus} Z$ & $Z$         & $T$    & $G$        & $T$             & boundary/edge \\[3pt] 
            & $T{\setminus} Z$ & $T$    & $T$        & $T$             & boundary \\[3pt]
            & $G{\setminus} T$ & $T$    & $c\,Tc^{-1}$ & $Z$             & interior
\end{tabular}
\end{center}
Note that the choice of the edges was arbitrary, 
so they must indeed have the same stabilizer as the rest of the boundary of $Q$
(except for the corners). 
The corners are the most singular points,
and have the largest stabilizer.

\section{The projected conjugacy classes}
In the large level limit, the D-branes in the coset $Q$ are localized at sets 
$\pi_{\rm Ad}\left(C_{abc}\right)$
defined as follows, cf.\ \cite{FW} for further notation and references. 
First, introduce the following notation for the conjugacy classes,\\[-6mm]
\be C_x:= \left\{ g(\psi,\theta,\phi) \in G | ~ \psi = x\right\}\,.
\ee
Second, define the operation $*$ on pairs of conjugacy classes,
\be C_x * C_y := \left\{ g=g_x g_y \in G | ~ g_x \in C_x\, , ~ g_y \in C_y \right\} \,.
\ee
The set $C_x * C_y$ is invariant under Ad$(G)$.  
Furthermore, 
introduce the sets
\be C_{abc} &: =& (C_a \times C_b) * C_c \\ \nonumber 
&:=& 
\left\{ (g_a g_c,g_bg_c) \in G\times G | 
~ g_a \in C_a\, , ~ g_b \in C_b\,,~ g_c \in C_c \right\} 
\, \subset \, G \times G \,.
\ee
These sets are the ones that project to the coset branes. 
Note that they are fixed under the adjoint action;\\[-9mm] 
\be  (g,g)\, C_{abc}\, (g^{-1},g^{-1}) = C_{abc}\,. \ee
By studying the sets $C_{abc}$, one can draw conclusions about the dimensionalities 
and other properties of the branes in $Q$. 
This is the topic of the rest of this note. First an observation.
The simplest example of a set which is invariant under the diagonal Ad-action is the 
orbit of a point $(p_1,p_2) \in  G\times G$. If this is a generic point (projecting
to the bulk of $Q$), the stabilizer is $\zet_2$, thus the orbit Ad$_{G}(p_1,p_2)$ is
three-dimensional in $G\times G$. The geometric picture of this 
orbit is that $(\psi_1,\psi_2)$ are fixed, $(\phi_1,\phi_2)$ take all values and 
we have a line in the $(\theta_1,\theta_2)$-plane.
If instead $(p_1,p_2) \in  Z\times Z$, the orbit Ad$_{G}(p_1,p_2)$ is zero- dimensional,
and projects to a corner in $Q$. Thus, we shall need not only the dimensionalities of 
$C_{abc}$ to find the dimensionalities of $\pi_{\rm Ad}\left(C_{abc}\right)$,
we also need the information about the stabilizers.

\paragraph{Corners} From \erf{inverse} we learn that 
conjugacy classes $C_x\subset SU(2)$ equal their own inverses; if $g\in C$, 
then $g^{-1}\in C_x$. For the $\psi_1=0$ edge to be contained in $C_{abc}$
it is necessary that $a=c$ because otherwise, $C_{c}$ does not contain the 
inverse of $C_{a}$. The equation \erf{inverse} tells us that this condition also is sufficient.
Thus, the {edge} defined by $\psi_1=0$ is reached 
if and only if $a=c$. For reaching the edge $\psi_1=\pi$, we ask: Given $g\in C_a$, 
we would like to have a $g'\in C_c$ such that $g(g')^{-1}=-1$. 
This is equivalent to that $C_c$ contains the inverses of $(-1)C_a= C_{\pi-a}$. But then,
$C_c= C_{\pi-a}$.
As argued above, the edges per se are not interesting, 
but it is useful to note that a corner is where two edges meet 
(now viewing $Q$ as a pillow, rather than a pinched cylinder). 
The analysis described above leads to the following conclusions
about the four corners (whose $\theta$-coordinates are irrelevant). \\[-2mm]

\noindent
$\cdot$ $(0,0)\in \pi_{\rm Ad}\big(C_{abc}\big)$ 
~ if and only if~ $a=b=c$.\\[3pt]
$\cdot$ $(0,\pi)\in \pi_{\rm Ad}\big(C_{abc}\big)$ 
~  if and only if~ ($a=c$ and $b=\pi-c$).\\[3pt]
$\cdot$ $(\pi,0)\in \pi_{\rm Ad}\big(C_{abc}\big)$ 
~ if and only if~ ($a=\pi-c$ and $b=c$).\\[3pt]
$\cdot$ $(\pi,\pi)\in \pi_{\rm Ad}\big(C_{abc}\big)$ 
~  if and only if~ ($a=\pi-c$ and $b=\pi-c$).\\[-2mm]

In particular, if a brane contains two corners, it automatically contains all four.
The fixed point brane (as mentioned in the introduction) is 
$\pi_{\rm Ad}\left(C_{fff}\right)$, with $f= \pi/2$.  
This is the unique brane that 
contains all four corners, and is therefore in some sense the most degenerate brane. 
We shall show that the fixed point brane fills up all of $Q$.
However, it is not the only brane that is three- dimensional. 

\paragraph{Dimensionality of the fractional branes}
It is instructive to first show that
\be C_f * C_f = G \,.
\label{lemma}\ee
Note
that \erf{lemma} is equivalent to
\be \forall ~ \,g \in G ~,~~~\,\exists ~ \,u,w \in C_f ~~: ~~ uw=g \,.
\ee
The product of two elements $u$ and $w$ on the equatorial conjugacy class
\be
C_f:=\left\{ g\in G~|~ \psi_g= \frac \pi 2   \right\} \,\subset G
\ee 
is computed using \erf{notation}.
\be uw= \bar n_u  \cdot \bar n_w\, \mathbf 1 
- \ii \,\bar \sigma \cdot (\bar n_u  \times \bar n_w)\,
\label{computation}
\ee 
Take an arbitrary element $g\in G$. The following shows that there exist
$u,w \in C_f$ such that $uw=g$.
Choose $\bar n_u\in S^2$ and $\bar n_w \in S^2$ in the plane perpendicular to $\bar n_g$. 
Choose the angle between 
them such that $\bar n_u\cdot \bar n_w = \cos \psi_g$, and so that 
$\bar n_u\times \bar n_w $ points in the same direction as $ \bar n_g$. Since the product 
$uw$ is a group element, we do not even need to check that 
$|\bar n_u\times \bar n_w | = \sin \psi_g$, and hence $uw=g$. 
This proves the claim \erf{lemma}.  

The following argument shows that 
\be C_{fff}= G\times G \,,
\label{spacefilling}
\ee
which implies that the corresponding brane in the coset is spacefilling, 
\be \pi_{\rm Ad}\left( C_{fff} \right) =Q\,.
\ee
First note that this is not an immediate consequence of \erf{lemma}, because the same 
factor $w \in C_f$ occurs at two places; we need to prove that for all $g_1,g_2 \in G$, 
we can choose $u,v,w \in C_f$ so that
$uw=g_1$ and $vw=g_2$. 
A quick glance at \erf{computation} 
reveals that the choice of $g_1$ forces $\bar n_w \perp \bar n_1$. 
Given $w$, one may choose 
$u\in C_f$ such that $uw=g_1$. Now it remains to prove that with $\bar n_w$ fixed in the plane  
$\perp \bar n_1$ through the origin, and for any $g_2\in G$, 
we can find $v\in C_f$ so that $vw=g_2$. 
We need $\bar n_w$ also to lie in the plane through the origin $\perp \bar n_2$. The two planes mentioned intersect at no less than two points on the sphere $C_f$. Choose $\bar n_w$ one of these points. Then choose $\bar n _v \perp \bar n_2$ such that $\bar n_w \cdot \bar n_v = \cos \psi_2$. 
It now follows $vw=g_2$. 
This proves the claim \erf{spacefilling}. 

\paragraph{Dimensionality of generic branes}
We proceed to determine the dimensions of $\pi_{ad}(C_{abf})$ for more 
general pairs $a,b$. The following argument shows that in the generic (nondegenerate) situation, 
that is when none of the labels $a,b$ takes 
value $0,\pi$, the set $C_{abf}$ is six- dimensional in $G\times G$, why the corresponding 
brane is three- dimensional in $Q$. 
One can parametrize conjugacy classes as 
\be u={\rm c}_ a ~ {\bf 1} + \ii \,{\rm s}_ a ~ \bar n_u \, 
{\times}\, \bar \sigma ~ \in ~ C_a\,,\ee
where ${\rm c}_ a\equiv \cos a$ and ${\rm s}_a\equiv \sin a$.  
With an analogous parametrization for $v\in C_b$, we need for $g_1=uw$ and $g_2=vw$ the following conditions,
\be \cos \psi _1  =  {\rm s}_ a ~ \bar n_u \cdot \bar n_w  ~~~~   && ~~~~
\sin \psi _1 \, \bar n _1 ={\rm c}_  a ~ \bar n_w + {\rm s}_ a~ \bar n_u \times \bar n_w \\ \nonumber
\cos \psi _2     = {\rm s}_ b~ \bar n_v \cdot \bar n_w  ~~~~ && ~~~~
\sin \psi _2 \,\bar n _2 = {\rm c}_  b ~\bar n_w + {\rm s}_ b ~\bar n_v \times \bar n_w \,.
\ee  
We immediately observe that in order for $(g_1,g_2)\in C_{abf}$, it is necessary that 
$\cos \psi_1 \leq |\sin a|$, and that $\cos \psi_2 \leq |\sin b|$. Assume that
$\cos \psi_1 < |\sin a|$, and that $\cos \psi_2 < |\sin b|$, which is possible because we required 
$a,b$ to be different from $0,\pi$. (This implies $\bar n_u \cdot \bar n_w \neq 1$ 
and $\bar n_v \cdot \bar n_w \neq 1$.) The rest of the argument aims to show that we now are in 
the interior of a 6- dimensional region.

Given allowed choice of angles $\psi_1,\psi_2$, we must keep the scalar products 
$\bar n_u \cdot \bar n_w $ and $\bar n_v \cdot \bar n_w $ fixed accordingly. 
We are still free to 
choose the vector $\bar n_w$ anywhere on $S^2$, and we are free to rotate the vectors 
$\bar n_u \times \bar n_w $ and $\bar n_v \times \bar n_w $ around the axis $\bar n_w$. 

Let us decide for a certain $\bar n_1$. With the freedom we have, we can certainly find 
vectors $\bar n_u$ and $\bar n_w$ such that $\cos a ~ \bar n_w 
+ \sin a~ \bar n_u \times \bar n_w $ points in the direction of $\bar n_1$. Since $uw$ is
a group element, it is automatic that 
$|\sin \psi _1 \bar n _1| = |\cos a ~ \bar n_w + \sin a~ \bar n_u \times \bar n_w |$. Thus, 
we require that the three vectors $\bar n_1$, $\bar n_w$ and  $\bar n_u \times \bar n_w$ 
lie in the same plane, and we fix the angle between them in this plane. We may rotate 
this plane, thus, we may rotate $\bar n_w$ around the
axis $\bar n_1$. Furthermore, the angle between these vectors is different from zero, because 
we assumed $\bar n_u \cdot \bar n_w \neq 1$, which implies $\bar n_u \times \bar n_w \neq 0$. 

As we rotate the above mentioned plane, 
$\bar n_w$ sweeps out a (nondegenerate) circle, around each choice of $\bar n_w$, the vector
$\bar n_v\times \bar n_w$ sweeps out another nondegenerate circle. Thus, the vector $\bar n_2$ is
allowed to reach a 2-dimensional subset of $S^2$. This completes the proof; subject to the 
constraints $\cos \psi_1 < |\sin a|$ and $\cos \psi_2 < |\sin b|$, $g_1=uw$ can take any value 
on the group, with enough freedom left to let $g_2$ sweep out a 3- dimensional subset of the 
group (varying $\psi_2$ and $\bar n_2$). Thus, the generic set $C_{abf}\subset G\times G$
is six-dimensional. \\[-2mm]

Let $a,b,c$ be an arbitrary nondegenerate triple of labels. 
These branes are 3- dimensional too. We have
\be g_1&=& ({\rm c}_  a\, {\rm c}_  c 
-{\rm s}_ a\, {\rm s}_ c \, \bar n_u \cdot \bar n_w ){\bf 1}
-\ii\, ( {\rm s}_ a\, {\rm s}_ c\, \bar n_u \times \bar n_w 
+ {\rm c}_  a\, {\rm s}_ c\, \bar n_w
+ {\rm c}_  c\, {\rm s}_ a\, \bar n_u)\cdot \bar \sigma
\nonumber \\
g_2&=& ({\rm c}_  b\, {\rm c}_  c-{\rm s}_ b\,{\rm s}_c \, \bar n_v \cdot \bar n_w ){\bf 1}
-\ii\, ({\rm s}_ b\, {\rm s}_ c\, \bar n_v \times \bar n_w 
+ {\rm c}_  b\, {\rm s}_ c\, \bar n_w+ {\rm c}_  c\, {\rm s}_ b\, \bar n_v)\cdot \bar \sigma \,.\nonumber
\ee
So assume $\psi_1,\psi_2$ are within the appropriate range, and fix the scalar products
$\bar n_u \cdot \bar n_w$ and $\bar n_v \cdot \bar n_w$ accordingly. The three vectors 
\be {\rm c}_  c \, {\rm s}_ a\,\bar n_u\,,~~ {\rm c}_  a\, {\rm s}_ c\,\bar n_w\,,
~~ {\rm s}_ a\,{\rm s}_c\,\bar n_u\times \bar n_w
\ee
span a polyhedron, whose one corner is at the 
origin, and the opposite corner is at $\bar n_1$. We may rotate this polyhedron around these
two points, and the rest of the proof carries over analogously, showing that also these branes 
project to something three-dimensional in the coset $Q$.

\paragraph{Dimensionalities of degenerate branes}
In \cite{FFFS} and \cite{FW} it was argued that symmetry- preserving branes in 
WZW models are non- degenerate. Nevertheless, we shall for the remainder of
this note study the corresponding degenerate branes in $Q$.
The branes $\pi_{\rm Ad}\left(C_{abc} \right)$ where the second index $b$ is 
degenerate were studied in \cite{FS}. To interpret these branes as
branes in the Virasoro minimal model is not without complications. 
In order to do that, we need to fix one of the levels to the value $l=1$,
which forbids us to use results for the large level limit.
At small level, 
the branes are smeared around nondegenerate conjugacy classes \cite{FFFS}, \cite{FW}.
Nevertheless, let us take $b=0$ (the case $b=\pi$ is analogous). 
\be u&=& {\rm c}_  a \, {\bf 1}+\ii\, {\rm s}_ a \, \bar n_u \cdot \bar \sigma 
~~~~~~~~~~~~~~
v= {\bf 1} ~~~~~~~~~~~~~~
w= {\rm c}_ c \, {\bf 1}+\ii \,{\rm s}_ c \, \bar n_w \cdot \bar \sigma \nonumber \\
g_1&=& ({\rm c}_  a\, {\rm c}_  c-{\rm s}_ a\, {\rm s}_ c \, \bar n_u \cdot \bar n_w )
{\bf 1}
-\ii \,( {\rm s}_ a\,{\rm s}_ c\, \bar n_u \times \bar n_w 
+ {\rm c}_  a\,{\rm s}_ c\, \bar n_w+ {\rm c}_  c\, {\rm s}_ a\, \bar n_u)\cdot \bar \sigma
\nonumber \\ \nonumber 
g_2&=& {\rm c}_  c  {\bf 1}
-\ii \, {\rm s}_ c\,\bar n_w \cdot \bar \sigma \ee 
We immediately see that the branes are localized along the slice $\psi_2=c$. 
The last of the above equations fixes $\bar n_w= \bar n_2 $. 
As before, $\psi_1$ has a certain allowed range. 
For fixed $\psi_1$, note that the scalar product 
$\bar n_u \cdot \bar n_w$ is fixed. Thus, $\bar n_1$ takes a one- dimensional subset of values 
on $S^2$. To sum up, the set $C_{a0c}$ is four-dimensional. If we project out the coordinates
$\theta_2$ and $\phi_2$ on the set $C_{a0c}$, then $\bar n _1$ takes values in 
a circle. Since we need to project out $\phi_1$, the coordinate $\theta_1$ is fixed. Thus, 
these branes $\pi_{\rm Ad}\left(C_{a0c} \right)$ are at most one-dimensional, 
extending only in the coordinate $\psi_1$.\\

If we also take $a=0$, $g_1= {\rm c}_  c  {\bf 1}
-\ii \,{\rm s}_ c\,\bar n_w \cdot \bar \sigma $, 
also $\psi_1$ is fixed; $\psi_1 = c = \psi_2$. 
Also note that $\bar n_1 =\bar n_w = \bar n_2$.
Thus, $C_{00c} \subset G\times G$ is two-dimensional and must project to a region where the
stabilizer is at least one-dimensional, that is at the boundary of $Q$, which is at 
$\theta = 0,\pi$. Thus, the branes $\pi_{\rm Ad}\left(C_{00c} \right)$ are pointlike.\\[-2mm]

If we consider $C_{a00}$, we have 
$g_1= {\bf 1} {\rm c}_  a -\ii\,  {\rm s}_ a\, \bar n_u \cdot \bar \sigma$, 
$g_2= {\bf 1}$.
This projects to a point $\psi_1 =a$, $\psi_2=0$ on the edge of $Q$.\\[-2mm]

The case left to consider
is $C_{ab0}$. We have $g_1= {\rm c}_ a\, {\bf 1}
-\ii \,{\rm s}_ a\, \bar n_u \cdot \bar \sigma$, and 
$g_2= {\rm c}_ b\, {\bf 1}
-\ii\, {\rm s}_ b\, \bar n_v\cdot \bar \sigma $.
Thus, $C_{ab0}$ is four-dimensional in the covering, and projects to a line that meets the 
boundary of the coset at both ends $\theta=0,\pi$.

\section{Conclusions and outlook} 
In the non- simply connected Lie group $SO(3)=SU(2)/\zet_2$, 
there are two fractional branes supported at the unorientable conjugacy class 
$\mathbb R\mathbb P^2\subset SO(3)$. The fundamental group of this conjugacy class is $\zet_2$, and the two branes correspond to different choices of gerbe modules \cite{GawG}.
In the CFT description, 
these fractional branes correspond to boundary states that are labelled by the fixed point of the simple current group $\zet_2$ of the $SU(2)$ WZW model.
In the model \erf{pillow} that was investigated here, 
we have a comparable situation: 
The CFT description at levels $k,l,k+l$ involves a 
$\zet_2$ identification group, with the generator 
acting as 
\be (a,b,c) \mapsto (k{-}a,l{-}b,k{+}l{-}c)\,.
\ee
In this paper, the fixed point boundary states have been shown to be supported at the whole target.
Other branes reach at most one of the singular points (corners).      
If there is a corresponding gerbe (module) structure for cosets, 
it seems to be unknown in the literature.  
Hopefully, the present investigations can be used as a guideline for
future work in this direction. 

\paragraph{Acknowledgements} The author would like to thank J. Fuchs and C. Schweigert
for helpful suggestions.

\end{document}